\documentclass[aps,prl,twocolumn,groupedaddress,showpacs]{revtex4-1}
\usepackage{amsmath}
\usepackage{graphicx}
\usepackage{epstopdf}
\usepackage{epsfig}
\usepackage{amssymb}
\usepackage{bm}
\renewcommand{\thefootnote}{\fnsymbol{footnote}}
\newcommand{\beq}{\begin{equation}}
\newcommand{\eeq}{\end{equation}}

\usepackage{color}

\begin{document}

\title{II. Non-commuting Matrix Solution of DGLAP; ${F_2}^{p,d}$ Data Leading to Partons Directly without Parameterization}
\pacs{12.38.Bx} 
 \author{M. Goshtasbpour$^{1}$,%\email{Goshtasb@sbu.ac.ir} 
and M. Zandi$^{2}$ \linebreak}
\affiliation{$^{1}$ Dept. of Physics, Shahid Beheshti University, G. C., Evin 19834, Tehran, Iran. \\ $^{2}$ II. Physikalisches Institut Georg August Universität Göttingen, Germany.} 

\begin{abstract}
 
 Dominant present path for determination of parton distribution functions (pdfs) from data is based on pre-assumed form of parametric pdfs. Here, an alternative direct, or non-parametric method of pdf extraction is spelled out. As the main task, least square estimates of the central values of pdfs are obtained at a chosen ‎$ Q^2 $, and at ‎$x_i, i=1,..., n$ of the analyzed ${{F_2}^{p,d}}$ data points. 
In the process,  numerically singular system of LO PQCD weighted linear combination of  decomposition equations of the data points, each at a given $(x_i, {Q_{ij}}^2), j=1, ..., n_i$, obtained from a respective $\chi^2$,  together with the equations of ZM VFN constraints, are solved. In each data equation, the corresponding data points are decomposed into their pdf components, evolved from the set of unknown pdfs at $(x_i, Q^2), i=1, ..., n$. A similar evolution is done in the constraints. As a complementary task, the constrained discretization of Bjorken $x$, required for 
 the commuting solution of DGLAP, \cite{I}, is relaxed, and 
 a non-commuting solution on the more natural set of exact $x$-points of the data
 is developed.  \\
\end{abstract}
 
\maketitle

\section{Introduction}

Decomposition equation of each data point serves as a weighted component of data equations on parton distribution functions (pdfs), and each ZM VFN constraint is a theoretical equation on pdfs, at different ‎$ Q^2 $ of the data and constraints, at a given order in PQCD. The integro-differential DGLAP evolution equations of the same order in perturbation may then be solved and used to bring all the pdfs to the unique ‎$ Q^2 $ of the unknowns in the resulting numerically singular system of linear equations to be solved (via SVD). Here, in the second paper of the series, the process is realized, for the simplest example, on ${F_2}^{p,d}$ data at leading order (LO), with possible wider implications% for phenomen
. 
‎
\section{non-commuting solutions of all DGLAP equations}
In contrast to commuting solutions of DGLAP equations, \cite{I}, with commuting banded (bdd) lower triangular (l.t.) splitting function matrices, here, more general non-commuting solutions of DGLAP equations, with non-commuting, non-bdd l.t. splitting function matrices are considered because of their advantage in data analysis.

\subsection{%\textit
${n}$ Dimensional $x$-Space of Data Points and each pdf Set}
Refering to the corresponding section in \cite{I},
 commuting solutions of DGLAP constrain the discrete $x$-basis, for the linear algebra, to be of the form 
 
\begin{eqnarray}
\label{2-1}
x_i=x_1^i, \hspace{5mm} i=1,\cdots ,n,
\end{eqnarray}

in order to have commuting banded (bdd) lower triangular (l.t.) splitting function matrices.
  Discrete $x$ in (\ref{2-1})  are not the exact $x$-points of the data. So, eventually, for data analysis, interpolation techniques should be used.

The non-commuting numerical solutions, to be presented in this section, can simply use the exact Bjorken $x={\lbrace{x_i}, i = 1,...,n\rbrace}$ ‎of the data set. On either discrete $x$-set, a discrete set of ‎\textit{n}‎ basis vectors for each pdf set is defineable. Thus, the splitting functions operating on this basis (space) can be calculated as ‎$ n \times n $‎ matrices.\\

\subsection{Flavor Evolution Equation}

At LO, where equality of splitting functions (neglecting quark masses), leaves only four different ones, the evolution equation can be written in the simple ‎form of $ (m+1) $‎ coupled equations of ‎$ m $‎ independent quarks and antiquarks coupled through gluon distribution, (\ref{1-5}). 
\begin{eqnarray}
\begin{gathered}
‎\frac{\partial}{\partial t}‎\begin{pmatrix}
q_1(x,t)\\
‎\vdots‎\\
q_{m}(x,t)\\
g(x,t)
\end{pmatrix}=\hfill \\ \hfill \\ \int_{x}^{1}‎\frac{dy}{y}‎‎\begin{pmatrix}
P_{qq}(\frac{x}{y}) & 0 & ‎\cdots‎ & 0 & P_{qg}(\frac{x}{y})\\
0 & \ddots & \ddots & ‎\vdots‎ & ‎\vdots‎ \\
‎\vdots & \ddots & \ddots & ‎0‎ & ‎\vdots‎ \\
0 & \cdots & 0 & P_{qq}(\frac{x}{y})‎ & P_{qg}(\frac{x}{y})‎ \\
P_{gq}(\frac{x}{y}) & ‎\cdots‎ & ‎\cdots‎ & P_{gq}(\frac{x}{y})‎ & P_{gg}‎(\frac{x}{y})
\end{pmatrix}. ‎\begin{pmatrix}
q_1(y,t)\\
‎\vdots‎\\
q_{(m)}(y,t)\\
g(y,t)
\end{pmatrix}
\end{gathered}
‎\label{1-5}‎
\end{eqnarray}

 where
\begin{equation}\label{1-3}
 ‎\frac{dt}{d Ln(Q^2)}=‎\frac{\alpha‎_s‎(Q^2)}{2\pi}.‎
\end{equation}

For $ n_f $ flavors,  
However, usually,  a linear combination, may reduce $m$ and modify coefficient of $ P_{qg}$, e.g., $m\rightarrow 2n_f -2$ for FIG. 1 of \cite{I}, 
  for use of total $c$ and $b$ pdfs; or $m \rightarrow n_f$ in the $F_2^{p,d}$ data analysis, in the present paper, due to use of $q_{total}$ for all the flavors. 

\textit{Flavor} evolution DGLAP (\ref{1-5}) is considered as an alternative to \textit{non-singlet, singlet ($NS - S$)} division of DGLAP already discussed in \cite{I}. Thus, both have the same number of independent parton distributions. Here, FIG. \ref{1} to  FIG. \ref{6} show the $(n_F +1)$ resulting pdfs of our $F_2^{p,d}$ data analysis.

Similar to the case of \textit{singlet}, \cite{I}, as all the kernels are $t$ independent, we expect the solution for finite $Q^2$ interval to be of the form:
\begin{equation}\label{3-8}
{f^F}(t) = {e^{(t - {t_0}){P_F}}}{f^F}({t_0}),
\end{equation}
where ${f^F},$ is a $(m+1)$ column vector of quarks and gluon and ${P_F}$ is the kernel of (\ref{1-5}) respectively. 

The \textit{singlet} and the \textit{flavor} kernels have some essential similarities that help in finding the analytic commuting solution of the \textit{flavor} DGLAP for finite $Q^2$ interval. 
The result may be  
presented in a future paper. For the present, we can have the numerical non-commuting solutions of  FIG. \ref{1} to  FIG. \ref{6}.

  ‎\subsection{Numerical Non-commuting Solutions  for Finite $Q^2$ Interval for all forms of DGLAP Equations}

Here, based on the exact Bjorken ‎$ x_i $ of data points, e.g. for the structure functions, numerical non-commuting solution of the DGLAP equations is presented.
DGLAP equations are of the generic form:
\begin{eqnarray}
\frac{{d{f_K}(x,t)}}{{dt}} = \int_x^1 {\frac{{dy}}{y}}{P^K}(\frac{x}{y}){f_K}(y,t),
\label{eq.14}‎
\end{eqnarray}
where $K$ can be ${NS}, S,$ or $ F$.
Having the matrix forms of the splitting functions kernels, section (2), matrix forms of the DGLAP equations are:
\begin{eqnarray}
\frac{{df_i^K(t)}}{{dt}} = \sum\limits_{j = 1}^i {{{P^K}_{ij}}f_j^K(t) \Leftrightarrow {\frac{d ‎\textbf{f}‎^K(t)}{{dt}}} = ‎{\textbf{P}^K}‎.{‎\textbf{f}‎^K}(t)}.
\label{eq.15}‎
\end{eqnarray}
Independence of the kernels from the variable $t$, leads the following solution to equation (‎\ref{eq.15}‎ ), for the finite evolution from $t_0$ to $t$.
\begin{eqnarray}
{‎\textbf{f}‎^K}(t) = {e^{‎(t - {t_0})‎\textbf{P}^K}}.{‎\textbf{f}‎^K}({t_0}) \equiv {‎\textbf{E}^{K}‎(t - {t_0})}.{‎\textbf{f}‎^K}({t_0}).
\label{eq.16}‎
\end{eqnarray}
The final $(t-t_0)$ dependent $‎\textbf{E}‎^K$ matrix constitutes the essential solution of DGLAP equations for finite evolution.

To determine the matrix form of $‎\textbf{E}^K‎(t - {t_0})=e^{‎{‎(t - {t_0})\textbf{P}^K}}$, the choice of (\ref{2-1})
 is mandatory to have commuting, banded triangular, splitting functions matrices,  for a finite expansion of the exponential in the analytical commuting solution of  (‎\ref{eq.16}), ‎\cite{I, Spin2000}. But experimental data points do not exactly match the ‎$ x $‎ points of (\ref{2-1}).  So, eventually, interpolation techniques should be used.
 
However, as hinted below equation (\ref{2-1}), such restrictions is not necessary. Exponential of the non-banded (non-commuting) form of the LO kernels $P_{NS}$, $P_S$ and $P_F$ can be numerically calculated, for the non-commuting solution of (‎\ref{eq.16}), at least within our limited data analysis. Using numerical algorithms to approximate the exponential in (‎\ref{eq.16}) brings freedom from banded triangular matrices, so one could use experimental ‎$x$ points directly to discretize the ‎$x$‎ space.

In this paper, we go directly to the non-commutative solutions,  based on the sequence of
 $x$ points of data, aimed at ${F_2}^{p,d}$ data analysis for finding   ${NS}, S,$ and $ g$ pdfs.   
The question of comparison  with flavor solutions can (may?) be trivial, considering prevailing linearity, or (massless) $SU(n_F), \ n_F\leqslant 5$ symmetry, or our ZM VFN constraints, discussed in the next section. 
It is also computationally tested via indirect flavor evolution of \cite{I}.
 
\section{Intoduction to data analysis of structure functions ${F_2}^{p,d}$} 

 LO decomposition equation (factorization) of ${F_2}$ is: 

\begin{equation}\label{3-1}
{F_2} = x(1 + \frac{{{\alpha _s}}}{{2\pi }}{C_q}) \otimes {\hat F_2^{q}} + x\frac{{{\alpha _s}}}{{2\pi }}{C_g} \otimes (\sum\limits_F {e_F^2} )g,  \hfill 
 \end{equation}
 with the quark structure function in term of $n_F=5$  flavors $NS - S$, with usual definitions for \textit{p} and \textit{d} 
 \begin{eqnarray}\label{3-2}
\begin{gathered}
 {\hat F_2^{q/p}} = \frac{1}{{90}}(22\Sigma  + 3{q^{(24)}} - 5{q^{(15)}} + 5{q^{(8)}} + 15{q^{(3)}}), \\ 
  \\
{\hat F_2^{q/d}}= \frac{{\hat F_2^{q/p}} + {\hat F_2^{q/n}}}{2} = \frac{1}{{90}}(22\Sigma  + 3{q^{(24)}} - 5{q^{(15)}} + 5{q^{(8)}}). \\ 
 \end{gathered}
 \end{eqnarray}
The simplest choice we have made by writing (\ref{3-2}), unrealistically with $SU_F(5)$ symmetry, needs explanation.

 Within a VFN point of view, if it was not for bringing in some 
 minimal mass effects  
 of heavy quarks $c$ and $b$ 
 , QED DIS would be limited to  
 three light quarks. Why so? Because zero-mass assumption for the three lightest quarks is well justified, $m_q^2 <<Q^2, \ Q^2$ in the whole range of data and simulation. There is the possibility  of differentiating square of electric charge, $e_q^2$ at $\gamma*$ quark vertex in proton or deutron DIS, so $u_{total}$ stands apart.  Isospin symmetry differentiates $d_{total}$ and $s_{total}$. In other words, detection of electric charge squared weight of $q_{total}$ may leave us two quarks in form of a singlet  $q^{S}$, and a non-singlet $q^{NS}$,    
and isospin symmetry differentiates data and the set of variables $q^3\propto {F_2}^{(p-d)}$ from $q^{NS}$ which now becomes $q^{8}$, the $SU_F(3)$ octet. In this sense, could we say there is something of an $SU_F(3)$ symmetry? 

It is trivial that  ${F_2}^{p,d}$ have no information on the non-singlet valence quarks. %From ${F_2}^{p,d}$ data analysis, we find no way of draging out information on the valence degrees of freedom  as we will be doing for non-valence nonsinglet degrees of freedom. Here, the $NS$ DGLAP equations for valence quarks are never invoked or used. 
  Thus, pdfs are extracted from  ${F_2}^{p,d}$ as total: sea and valence are not separable. 

Bringing in minimal mass effects of heavy quarks, $c$ and $b$, to the measured data points, $F_2^{p}(x_{ij}, {Q_{j}}^2)$ and $F_2^{d}(x_{ij}, {Q_{j}}^2)$, in different regions of $Q^2$, corresponding to 4, and 5 flavors, allows unveiling of the $4$th and $5$th flavors.  $SU_F(3) \longrightarrow SU_F(5)$  symmetry under the assumption of having five zero-mass flavors symmetrically, as unrealistic or highly errored,  as a simple ZM VFN scheme may be. 
  
  Under $SU_F(5)$,   differentiation of $q^{15}$ and $q^{24}$ once again leaves  $q^{NS}=q^8$ as it was. Thus, 
two additional sets of nonsinglet degrees of freedom are brought in 
 $q_{n^2-1}$ with addition of the $nth$ quark as a heavy flavor as $Q^2$ increases. Finally, the number of extractable pdf central value sets increases to $n_F+1$.
  The additional one pdf is the gluon of gluon vertex.

Simulation of our direct method  is a testing ground for this point of view, its correction and refinement. 
Investigation of deviation from $SU_F(5)$  symmetry in the well-known regions of $Q^2$, is left out to be dealt with soon. 

 Minimal mass effects, at masses of $c$ and
  of $b$ quarks, for us, are continuity constraints $c(Q^2=m_c^2)=0$, and $b(Q^2=m_b^2)=0$   
  at Bjorken $x={\lbrace{x_i}, i = 1,...,n\rbrace}$ ‎of the data set, at the boundary between regions of $Q^2$ 
  where  it is assumed that $n_F$ respectively crosses $3$ to $4$, and  $4$ to $5$ flavors. In other words, constraints equate the singlet, $\Sigma$, and the $({n_F}^2-1)$-multiplet ($NS$) of the $SU(n_F)$ flavor symmetry of the region above each 
 boundary; thus, dropping the flavor number, $n_F$, by $1$ at and below the boundary. The constraint equations, together with consistent evolution, keep the continuity of the pdfs while the flavor numbers change at the boundary.  
 
Along with continuity  of pdfs, we have 
  continuity for the coupling constant as a changing $Q^2$ crosses massive quark $m_b^2$.
  
  Positions of $Q^2=m_b^2$ is in practice very different from $Q^2=m_c^2$. $Q^2=m_c^2$ lies asymptotically outside the $Q^2$ range, as equations of BCDMS data, \cite{BCDMS}, are simulated here.  It is only for $b$ that a break takes place when evolution between the data and our simulation 
 crosses $Q^2 =‎ m_b^2$.  
 Then, solving DGLAP for finite evolution $[Q_j^2, Q^2]$ is divided into two stages of $[Q_j^2, m_b^2]$ and $[m_b^2, Q^2]$. Furthermore, there are two sets of inputs, the first set has $n_f=4$ flavors at $Q_j^2$, the second set has $(n_f=5)$ flavors at $m_b^2$. In addition, the second set of inputs is the first set of outputs with an extra input pdf for the heavy quark, $b(Q^2=m_b^2)‎\equiv‎ 0$, our minimal effect of mass of $b$! In order to carry out the comparison with MSTW, we take their value, $m_b=4.75Gev$ \cite{MSTW}.

  In the direct, non-parametric, data analysis of DIS, a
  singular system of linear equations of data and VFN constraints is encountered. Direct pdf variables of this method are a few times more (next section), also in \cite{III}, 
   than the number of parameters of the parametric data analysis. 
A very simple idealization, a ZM-VFN scheme employed via the constraint equations, brought the first  successful removal of singularity via SVD. It is presented in its simplicity here, before further realistic elaborations of the effects of heavy quarks masses. In a coming step, %  \cite{IV}, 
 a critique, including that of the massless coefficient functions, (\ref{2-3}) and (\ref{2-5} ), having large errors, 
 is to be done. 
 
\subsection{Example Calculation of Coefficient Function Matrices}

The kernel of the convolutions, in DIS data analysis, is either the splitting functions for DGLAP, or the coefficient functions of the hadronic structure function, e.g. (\ref{3-1}). Construction of the matrix coefficient functions, for DIS, follows closely the path of construction of splitting functions, resulting in either commuting, banded lower triangular, or noncommuting matrices, respectively,  only depending upon whether equation (‎\ref{2-1}) is followed for discretization of $x$ or not. The two possibilities lead to analytical, \cite{I}, or more general numerical approachs. \\

 With examples on splitting functions, derivation of $P_{qq}$ was spelled out in \cite{I}. In a second example calculation, the matrix form of the LO ${\overline{MS}}$ coefficient functions  $ {C_q}(x) $  
 and $ {C_g}(x) $, suited for our ZM VFN scheme, are derived here, beginning with \cite{Stirling}:

 \begin{eqnarray}\label{2-3}
\begin{gathered}
 {C_q}(z) = \frac{4}{3}[2{(\frac{{\ln (1 - z)}}{{1 - z}})_ + } - \frac{3}{2}{(\frac{1}{{1 - z}})_ + } - (1 + z) \times \\
 \ln (1 - z) - \frac{{1 + {z^2}}}{{1 - z}}\ln z + 3 + 2z - (\frac{{{\pi ^2}}}{3} + \frac{9}{2})\delta (1 - z)]. \\ 
 \end{gathered}
  \end{eqnarray}

Using definition of "+" regularization, integrating $C \otimes {\hat F_2^q}$, (\ref{3-1}), by parts, in which  $d{v_1} = \frac{8}{3}\frac{{\ln (1 - y)}}{{1 - y}}dy$, and $d{v_2} = \frac{4}{3}[ - (1 + y)\ln (1 - y) - \frac{{1 + {y^2}}}{{1 - y}}\ln y + 3 + 2y]dy$; the matrix of coefficients of $n-$tuple $q$ may be read as:

  \begin{eqnarray}\label{2-4}
 C_q : \left\{\begin{gathered}
 (C_q)_{ii} = a + \frac{1}{x_i - x_{i - 1}}\int\limits_{x_i}^{x_{i - 1}} \{ v_1(\frac{x_i}{y}) + v_2(\frac{x_i}{y})\} dy  \\ 
 (C_q)_{ik} = \frac{1}{x_k - x_{k - 1}}\int\limits_{x_k}^{x_{k - 1}} dy\{ v_1 (\frac{x_i}{y}) +\\
  v_2 (\frac{x_i}{y})\}  -  \frac{1}{x_{k + 1} -x_k}\int\limits_{x_{k + 1}}^{x_k} dy\{ v_1(\frac{x_i}{y}) + v_2(\frac{x_i}{y})\} . \\ 
 \end{gathered}\right.
 \end{eqnarray}
 where $a={v_2}(1)-{v_1}(0) - {4{\pi ^2}}/{9}-6$. \\

Similarly for ‎$ {C_g}(x) $‎, beginning with
\cite{Stirling}‎: \\
 \begin{equation}\label{2-5}
{C_g}(z) = \frac{1}{2}[((1 - {z^2}) + {z^2})\ln (\frac{{1 - z}}{z}) - 8{z^2} + 8z - 1],
 \end{equation}
  
we will get the matrix form:
\begin{eqnarray}\label{2-6}
 {C_g} : \left\{ \begin{gathered}
 {({C_g})_{ii}} = v(1) + \frac{1}{{{x_i} - {x_{i - 1}}}}\int\limits_{{x_i}}^{{x_{i - 1}}} {v(\frac{x_i}{y})dy}  \\ 
 {({C_g})_{ik}} = \frac{1}{{{x_k} - {x_{k - 1}}}}\int\limits_{{x_k}}^{{x_{k - 1}}} v(\frac{x_i}{y})dy - \\
 \frac{1}{{{x_{k + 1}} - {x_k}}}\int\limits_{{x_{k + 1}}}^{{x_k}} v(\frac{x_i}{y})dy   \\ 
 \end{gathered} \right.
 \end{eqnarray}\\

where $dv ={C_g}(y)dy$. \\

\section{extraction of least square estimates of the central values of pdfs from ${F_2}^{p,d}$ data and their comparison with MSTW pdfs} 
At this stage, we are ready, with a bare minimum required for SVD to work, to get  the central values of pdfs from ${F_2}^{p,d}$ data. For the numerical solution, each ${F_2}^{p,d}$ data point at a given $(x_i, {Q_ij}^2)$ of an experiment, e.g BCDMS \cite{BCDMS}, is decomposed into its partonic components via (\ref{3-1}) and  (\ref{3-2}), evolved by our solutions of DGLAP, from a set of unknown pdfs, $u_k, k=1,...,m=(n_F+1)\times n-1$,\thefootnote{*}
at $(x_i, {Q}^2), i=1,...,n$, with a chosen ‎$ Q^2 $. 

The linear system of ${m}$ data equations and $(2n-1)$ ZM VFN constraints may be denoted as:
\begin{equation} \label{3-5}
A_{ij}{u_j}=b_i, \\ i=1,...,m+(2n-1), \\ j=1,...,m, \\ \\ \\  \hfill 
 \end{equation}
where the data equations come from minimization of ‎${\chi^2}$:
\begin{equation}\label{3-3}
{\chi ^2} = \sum\limits_{ij} {\frac{{{{(F_{2ij}^L - F_{2ij}^R)}^2}}}{{\sigma _{ij}^2}}},
 \end{equation}
with respect to the unknowns, 
\begin{equation}\label{3-4}
\frac{\partial {\chi^2}}{\partial {u_k}}=0, k=1, ..., m.
\end{equation} 
 ‎$ F_{2ij}^L $‎ is the value of data point on the left side of  (\ref{3-1}), with ‎ ${\sigma _{ij}}$ its quadratically calculated total error, and ‎$ F_{2ij}^R $‎ its LO decomposition corresponding to the right side of (\ref{3-1}, \ref{3-2}). In (\ref{3-3}), for data sets such as BCDMS \cite{BCDMS}, a sum over proton and deutron data points is understood. 
 
 The cut on data is set, in principle, to separate the $Q^2$regions of PQCD and higher twists at the invariant mass squared $W^2=20 Gev^2$, similar to MSTW LO cuts, \cite{MSTW, III}. Number of proton and deutron data points used are $153$ and $146$ respectively, thus less than ${4}\%$ of the total BCDMS \cite{BCDMS} data is cut out.  
  Ability to utilize correct cuts is a major improvement of the present version of the paper. It will be further discussed in \cite{III}. 

 Least square estimates of the central values of unknown pdfs $u_k, k=1,...,m$, at the chosen ‎$ Q^2 $, are obtained by solving the numerically singular system of linear 
   equations (\ref{3-5})   of the data, and ZM VFN constraints. Singular value decomposition (SVD) is the essential tool for bringing out the physically acceptable solutions from the context of singularity. In \cite{III} there will be extended exposition of the workings of SVD. Here, is how we learned to use SVD.
 
  Matrix of the coefficients $A$ in (\ref{3-5}) is singular.
 SVD helps separating and managing the singularity, namely the null subspace of the linear space of the singular matrix \cite{Press}. 
The magic of 
SVD here is to pinpoint numerically too small, deletable, eigenvalues, corresponding to a deletable set of eigenvectors of the numerical null subspace. 

 Operationally, i.e, in the process of trial motion up or down an indicator of the scale of ordered eigenvalues in the simulation program, 
 there is a single physical criterion for uncovering the border of null subspace, or the place (the indicator value) of its largest eigenvalue: 
 sudden appearance of well patterned, physically acceptable, set of solutions of the linear system, here the LS estimates of every pdf set, which takes place along with the deletion of the corresponding null subspace.\thefootnote{**}
 We'll be using this concept as the \textbf{first} characterization or "qualitative, intuitive, physical definition" of null subspace of a singular matrix, developed  quantitatively in \cite{III}.

\subsection{Results}

\begin{figure}[ht]
\includegraphics[height=.23\textheight, width=1.\columnwidth]{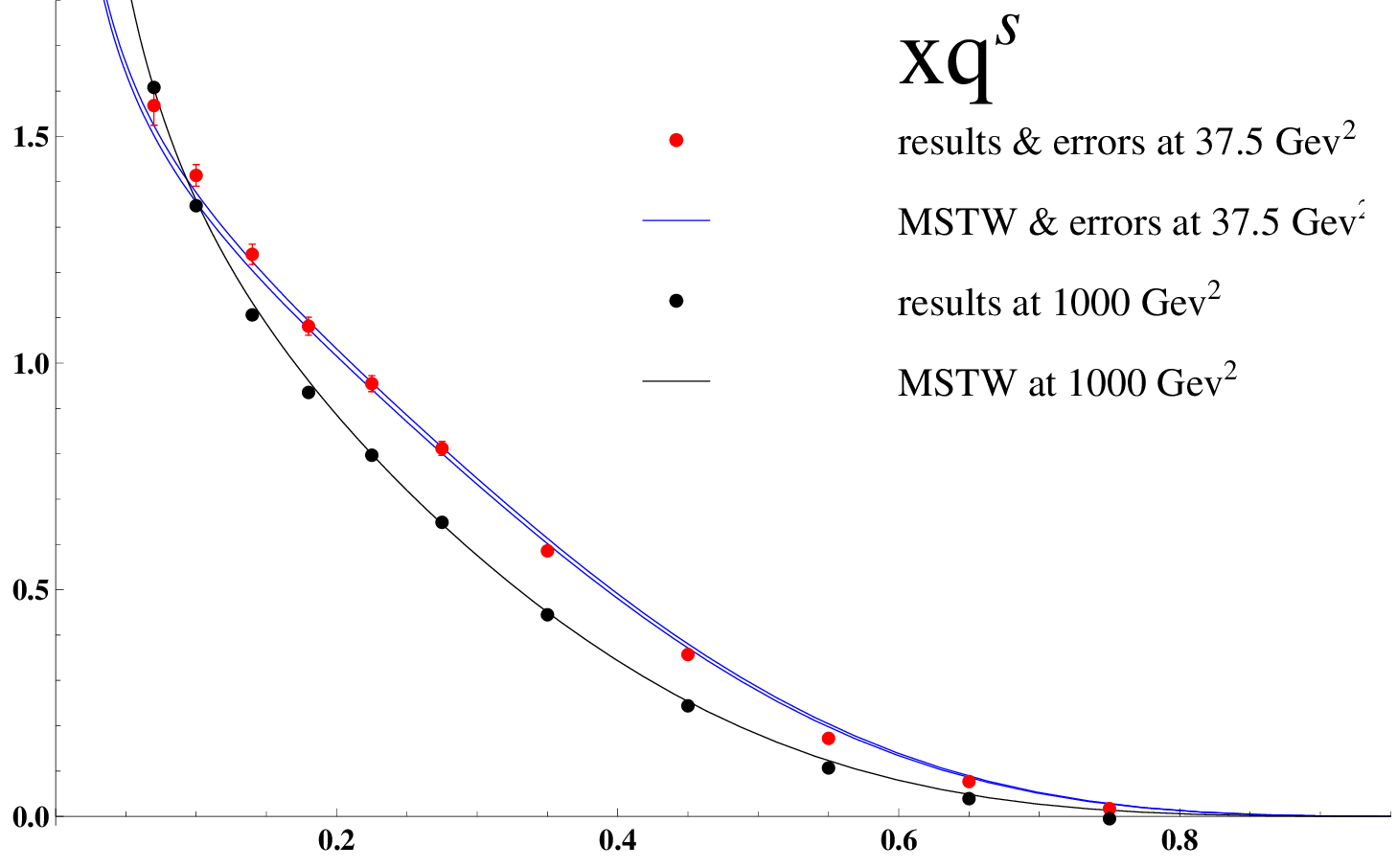}
\caption{Points of the SU(5) singlet pdf at the $x$-points of the BCDMS ${F_2}^{p,d}$ data from which they are extracted, in comparison with MSTW's. The choices of ${Q^2}=37.5$ and $1000 {Gev^2}$ of the graphs are arbitrary. The first is chosen near the center of population of the data points, in the non-asymptotic area of our ZM VFN in conflict with GM VFN of MSTW. The second is in the deep asymptotic area. The error analysis is done in \cite{III}. This is the most exact of the directly extracted pdf points from ${F_2}^{p,d}$ data point.
}
\label{1}
\end{figure}

\begin{figure}[ht]
\includegraphics[height=.23\textheight, width=1.\columnwidth]{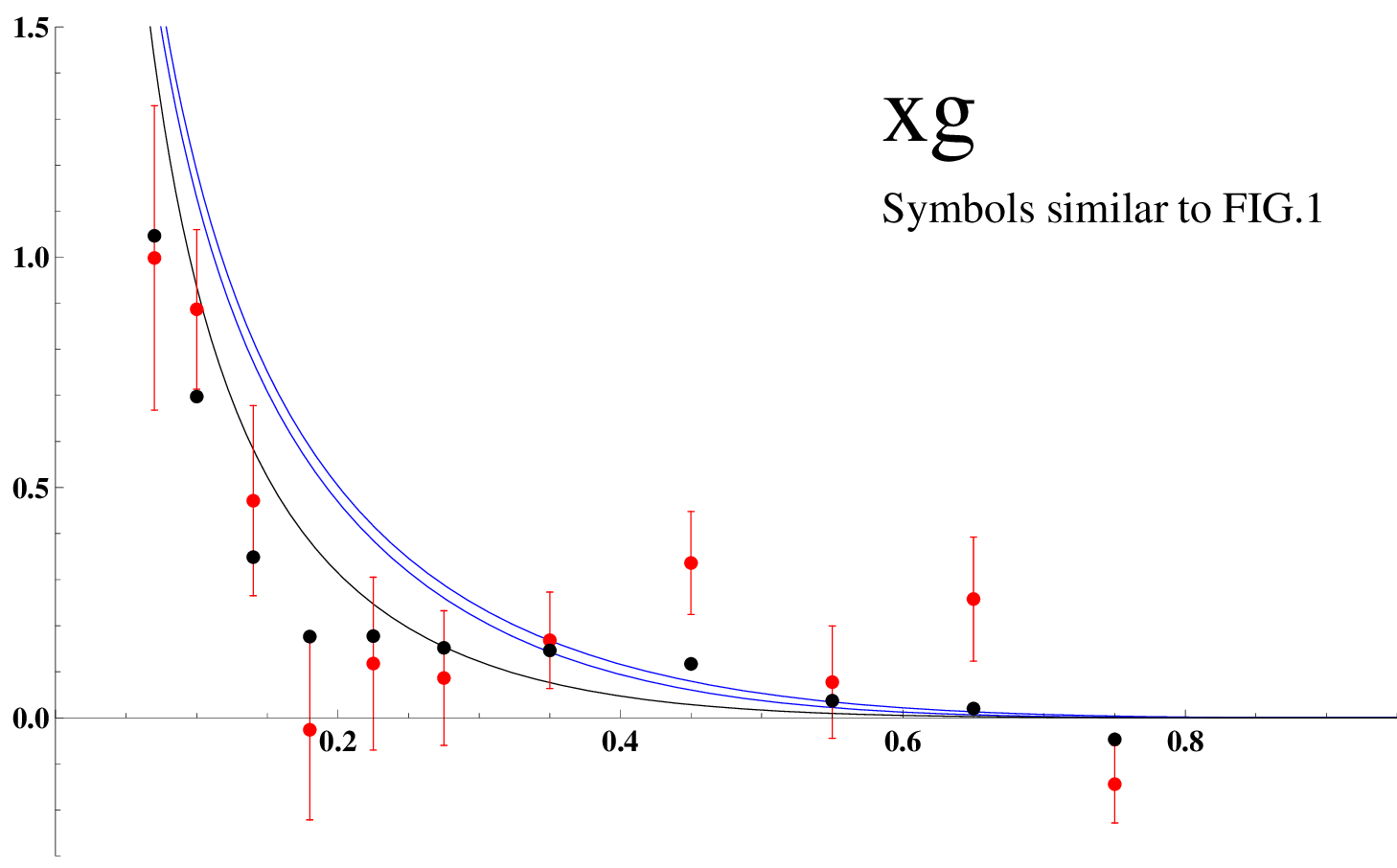}
\
\caption{Gluon; graphic symbols and caption similar to those of FIG. \ref{1}. This is the least exact of the extracted pdfs points. Partially, smoothened at ${Q^2}=1000 {Gev^2}$ due to larger intervals of evolution.
}
\label{2}
\end{figure}

The results are presented in the first six figures, with similar graphic symbols (described in the 1st legend), at two values of ${Q^2}=37.5$ and $1000 \ {Gev^2}$, at which solutions are computed independently, and then compared with MSTW. The first  $Q^2$ is chosen near the center of population of the data points, in the non-asymptotic area of our ZM VFN in conflict with GM VFN of MSTW; the second is in the deep asymptotic area. 
FIG. \ref{1} shows the SU(5) singlet, the most exact of the directly extracted pdf points from ${F_2}^{p,d}$ data point. FIG. \ref{2} shows the five flavor gluon, the least exact of the extracted pdfs points.

\begin{figure}[ht]
\includegraphics[height=.23\textheight, width=1.\columnwidth]{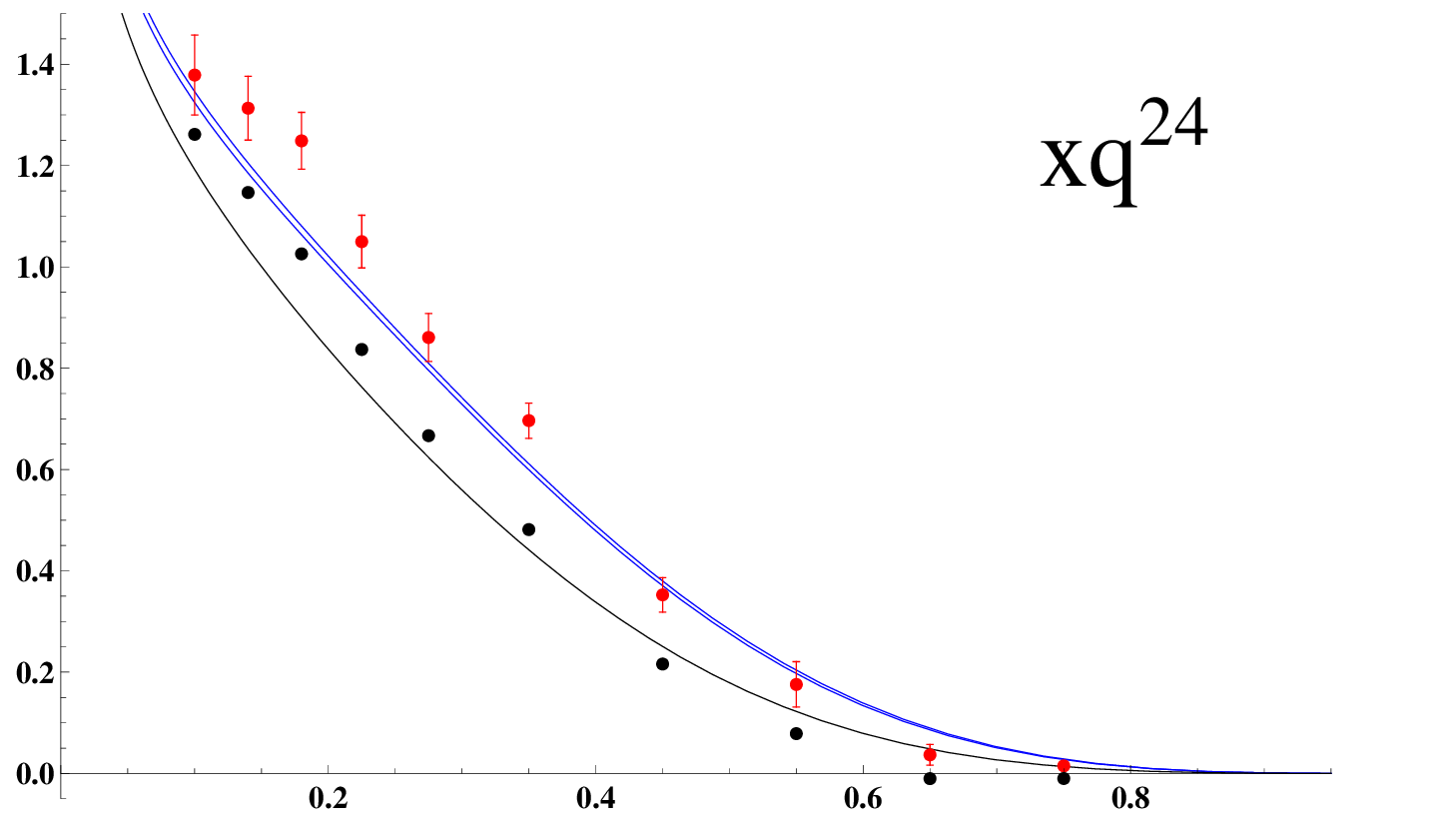}

\caption{ $NS$ pdf $xq^{24}$. Graphic symbols and caption similar to those of FIG. \ref{1}. There is an observable large-$x$ small-$x$ mismatch with MSTW as described in the text, common to all NS pdfs.
}
\label{3}
\end{figure}
\begin{figure}[ht]
\includegraphics[height=.23\textheight, width=1.\columnwidth]{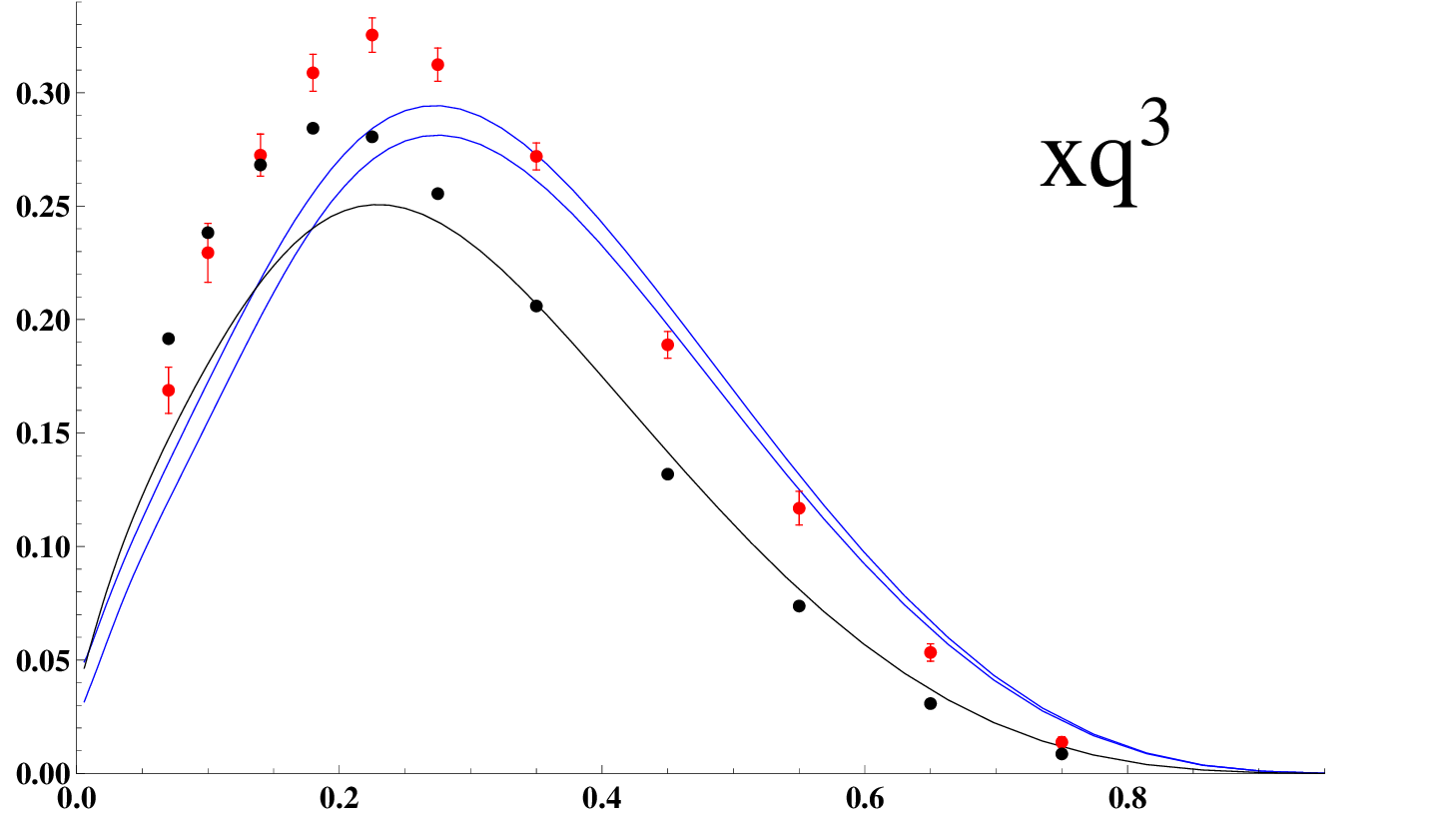}

\caption{ Independendent (in the sense of text in "Discussion")  $NS$ pdf $xq^{3}$. Caption similar to that of FIG. \ref{3}.}
\label{4}
\end{figure}
\begin{figure}[ht]
\includegraphics[height=.23\textheight, width=1.\columnwidth]{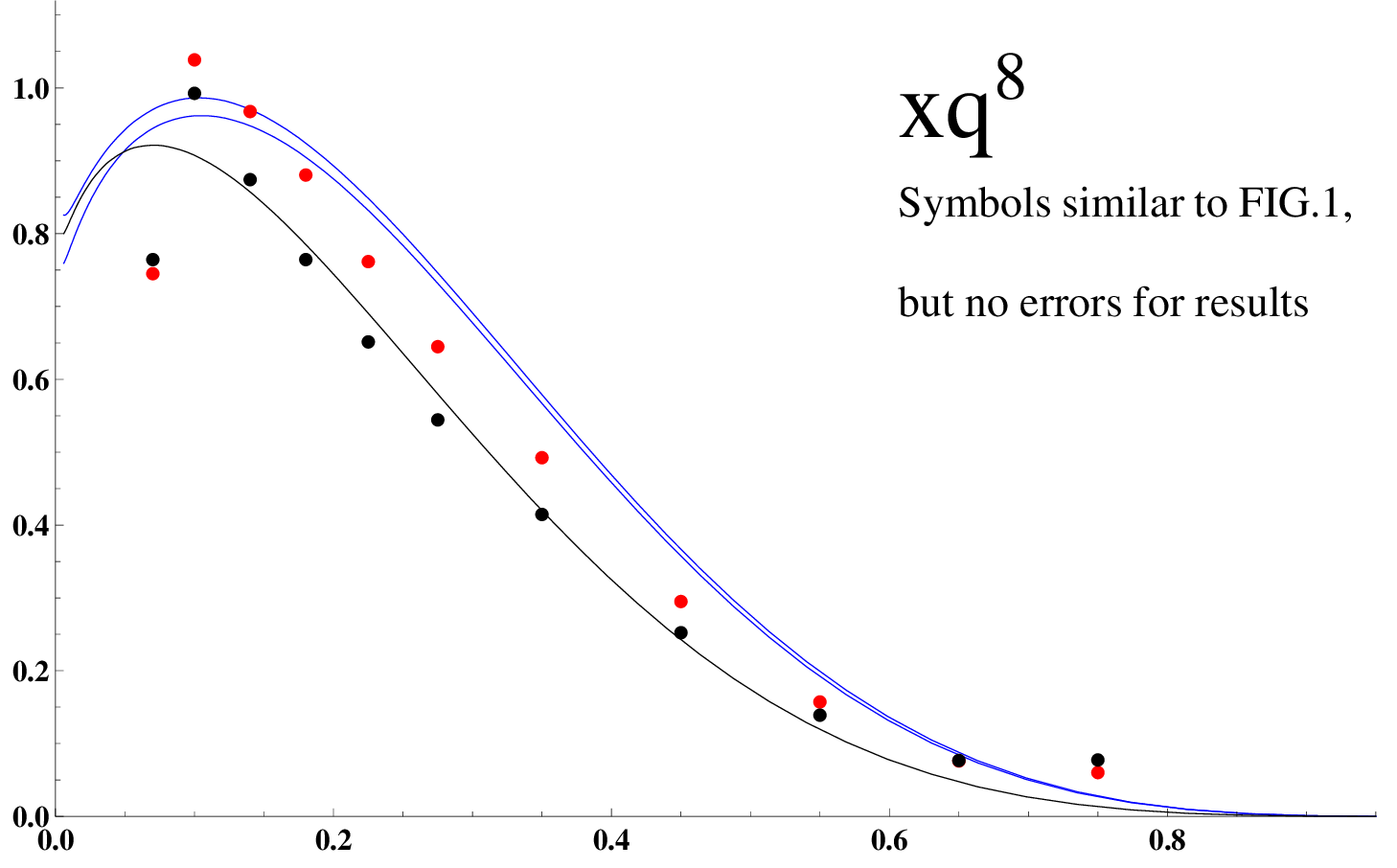}

\caption{$NS$ pdf $xq^{8}$. Caption similar to that of FIG. \ref{3}. Errors and their discussion is left to \cite{III}.}
\label{5}
\end{figure}
\begin{figure}[ht]
\includegraphics[height=.23\textheight, width=1.\columnwidth]{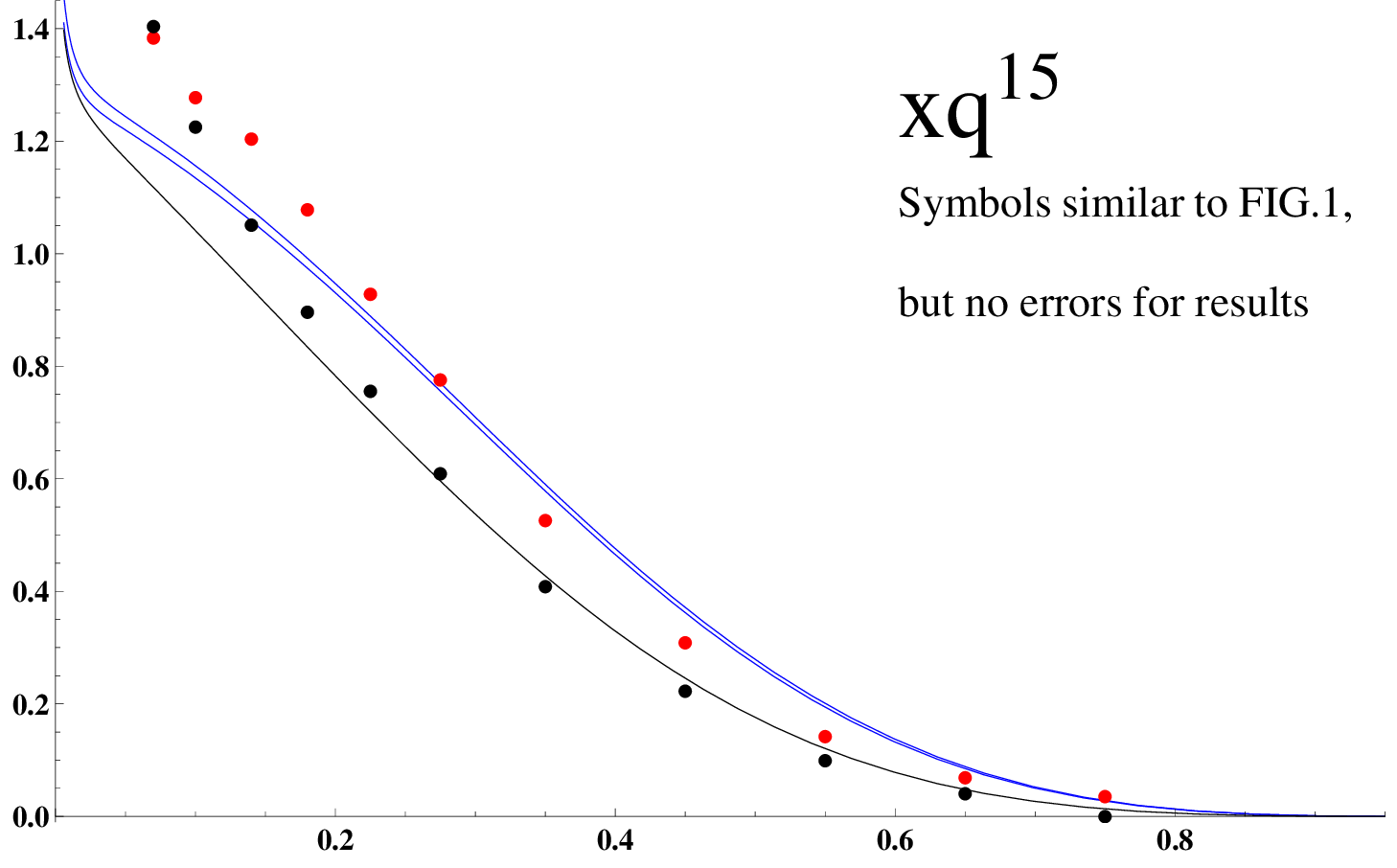}

\caption{$NS$ pdf $xq^{15}$. Caption similar to that of FIG. \ref{3}.  Errors and their discussion is left to \cite{III}.}
\label{6}
\end{figure}

Choosing ${Q^2}=1000$ well in the asymptotic region will not improve the obsevable lower large-$x$, and higher small-$x$ mismatch with respect to MSTW, but smoothens out the gluon and the $NS$ pdfs (except $q_3$ because of its independence) 
 due to larger intervals of evolution, taking place while setting up equations (\ref{3-4}). before obtaining the SVD solutions.

 \begin{figure}[ht]
\includegraphics[height=.23\textheight, width=1.\columnwidth]{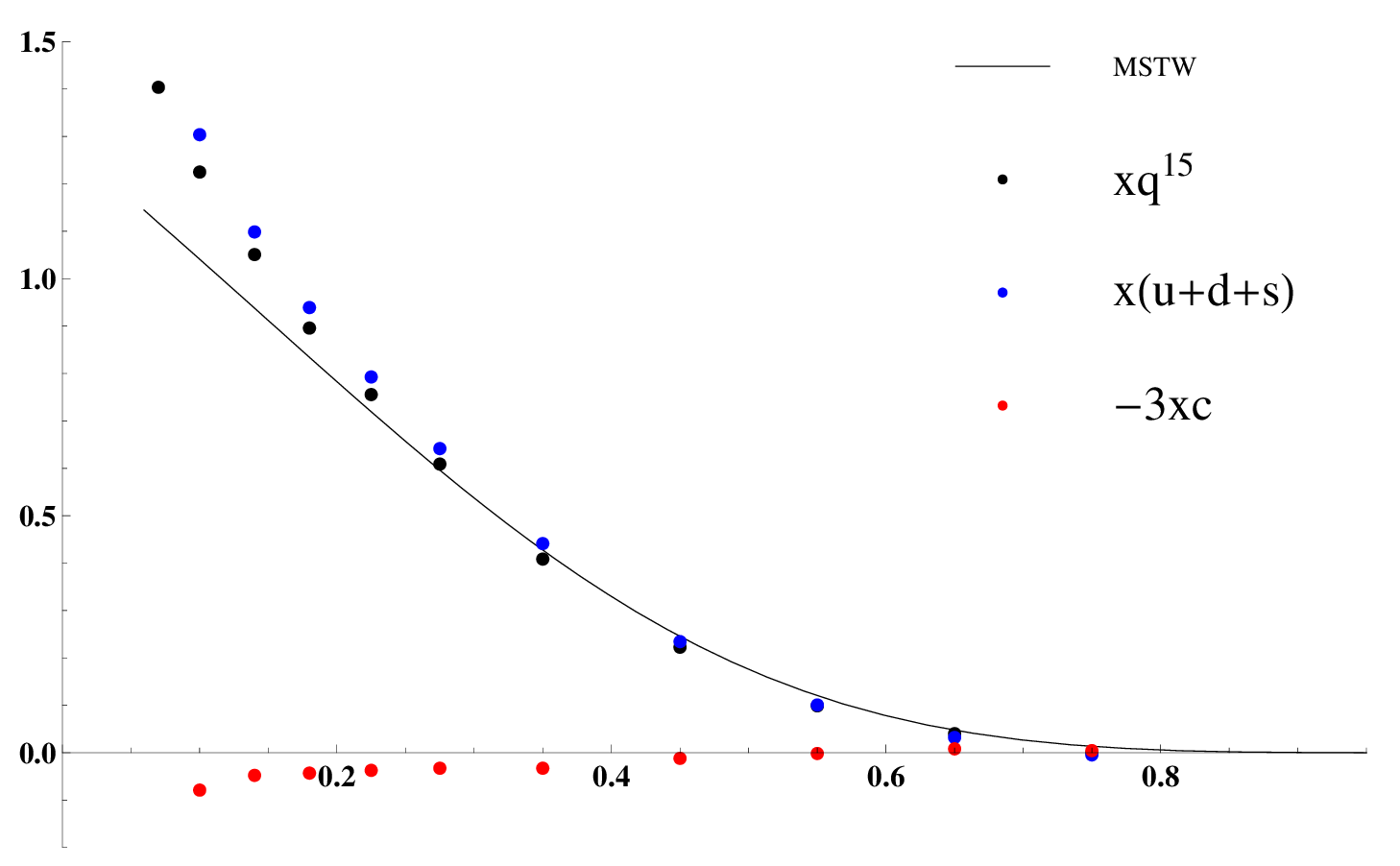}

\caption{At  ${Q^2}=1000 \ Gev^2$, MSTW (solid line), and our $NS$ pdf $q^{15}$ ( $11$ black dots) and its two, positive $(u+d+s)$ and negative $(-3c)$, components ($10$\thefootnote{*} dots in color each), with both series of  $c$ and $b$ constraints.  SVD produces acceptable physical results.}
\label{7}
\end{figure}

\begin{figure}[ht]
\includegraphics[height=.23\textheight, width=1.\columnwidth]{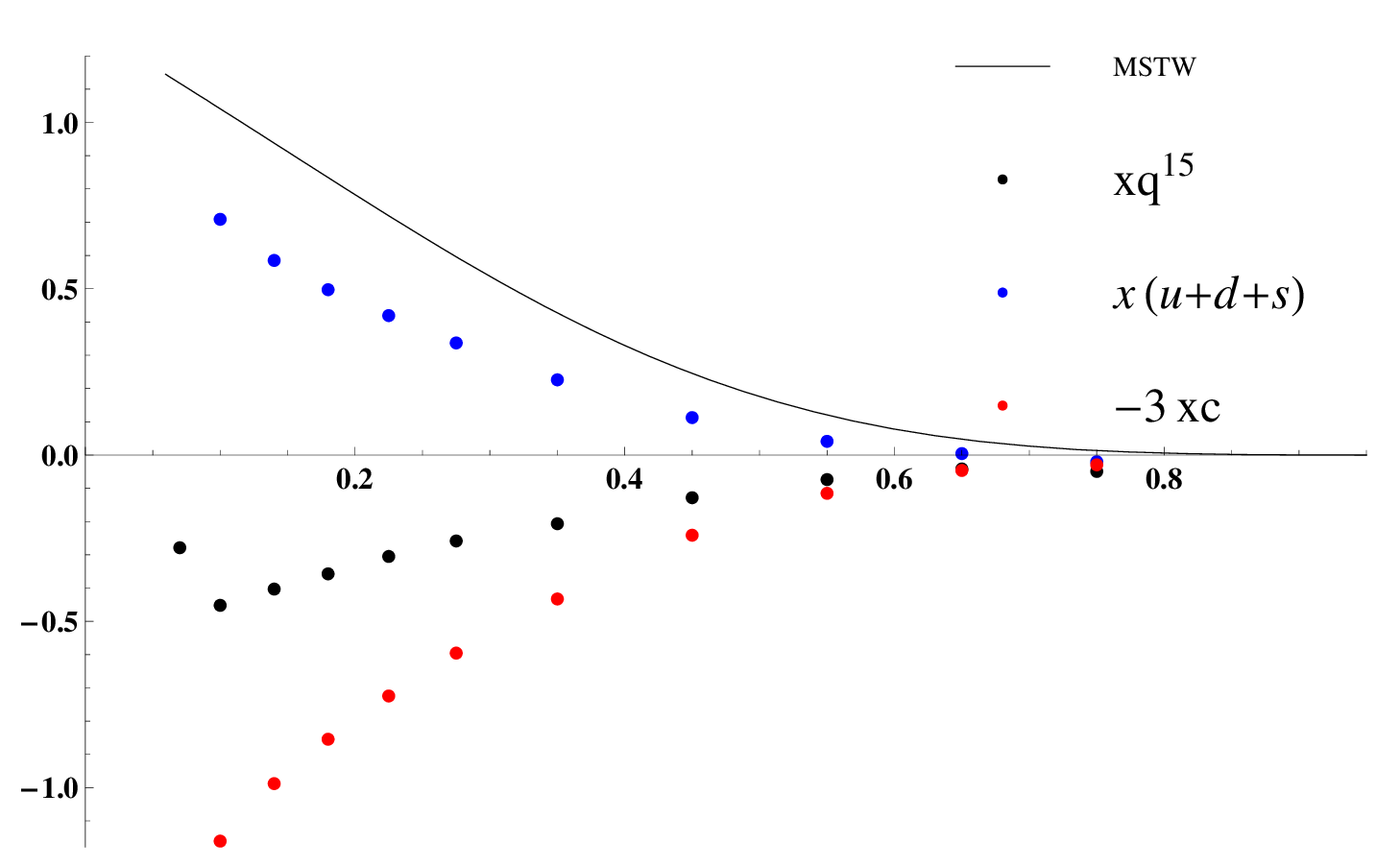}

\caption{Here, everything is the same as in FIG. \ref{7}, but without the series of $c$ constraints.  $q^{15}$ is the most adversely affected pdf from 
\textbf{lack of ZM VFN constraints at} ${Q^2}=m_{c}^2$.
 Here, SVD cannot produce physical results
.}
\label{8}
\end{figure}

\subsection{Discussion and prospects}
 The lower large-$x$, and higher small-$x$ mismatch with respect to MSTW, pronounced in all $NS$ pdfs, seems to come out of an x-dependent convolution found in factorizatins. Thus we may have a clue where to search for it.    \newpage     First of all, we intend to remedy 
 the very simple ZM VFN constraints (scheme), entering abruptly at the boundaries ${Q^2}=m_{c,b}^2$. This is to be done after the present paper.

Here attention is drawn to the following computational test of the theoretical necessity of havig our $c$ quark constraints as a possible minimum mass effect for having $(n_F + 1)$ pdf central values. 
 
At  $Q^2=m_c^2=1.96 \ Gev^2$, constraints may not appear to be %relevant
  as crutial as those at $Q^2=m_b^2$ from a practical point of view. We are in the asymptotic region for $c$ pdf, when working above the minimum of BCDMS data at ${Q^2}=7.5 \ Gev^2$. At $m_c^2$, even these constraints are overlapping with the higher twist boundary set at $W^2=Q^2(1/x-1)=20\ Gev^2$. However, even in that  asymptotic $Q^2$ region, 
   without $c$ constraints,  SVD cannot produce physical results,   FIG. \ref{7} and FIG. \ref{8}, magically (at this stage of understanding) pointing out a deep deficiency in such a practical point of view. 
  Anchorage of the pdf $c(m_c^2)=0$ sets essential physical constraints towards  asymptotic use of our ZM VFN, whether evolution takes place in the asymptotic region for $c$ pdf 
   or not.

 Operating. At, $m_b^2=22.5625 \ Gev^2$, in the middle of the finite $\Delta{Q^2}=[7.5, 230] \ Gev^2$ interval of BCDMS data, constraints are not only unavoidable in the above sense, but also are practically used in process of evolution through $Q^2=m_b^2$. 

The path towards global data analysis, begining with electroweak and going to hard scattering, to determine $(2n_F+1), n_F\rightarrow 6$ Standard Model pdfs is open. Quadratic non-linearity of hard scattering may need due attention.

 The path towards higher order analysis seems not only open via the numerical noncommutative solutions to DGLAP presented, %introduced in this paper,
  but also at least in some cases analytically ; e.g., %in the mentioned external product spaces involving l.t. bdd matrix algebra  of \cite{I},
   we have developed an NLO $NS$ commuting solution of DGLAP. Most importantly, A Higher Order Perturbative Parton Evolution Toolkit,  HOPPET, \cite{HOPPET}, which  we became familiar with only very recently, is exactly of the same family as ours, and of great value in this respect.

\section{Acknowledgements}
Thanks to  P. G. Ratcliffe for inception of the idea.

\thefootnote{*} Subtraction of one, from the number of unknowns and the corresponding constraint equations, is an indication of lack of information or data on a particular pdf variable, namely, $b(x_{11}=.07)$. In otherwords, to compare our results with MSTW, we use their constants including $m_b^2=22.56 \ Gev^2$, and at $x_{11}=.07$, BCDMS's highest $Q^2=19<22.56 \ Gev^2$. This leads to a missing point at $x_{11}=.07$ in graphs of the results wherever $b(x_{11}=.07)$ is involved.

 \thefootnote{**}% Rediscovery of this "operation" within the process of our work was due to MZ in fall of 2009 (MG: ??   Witholding, anomaly in my method  )
A note on the history of this point. The 3rd reference of \cite{Spin2000}, has a good overlap with this paper. There, we went as far  as decomposing $F_2$ of NMC \cite{NMC} with the Commutative Matrix solutions of DGLAP with the best fit of Bjorken $x$ given in Table I of \cite{I}, and discovering the singularity of resulting linear system of equations, an obstacle in the way of further development in 2002, \cite{Spin2002}. In some %the intervening
 years,  %before \cite{MZ thesis}
%In a long %isolated process, 
 application of the essential tool, SVD, %in MG's computational physics classes became well developed, notably in the notebook of ?. Salahshur , spring(?)2006
 was %found and
  learned; %\cite{site}
 trial and error for discovery in phenomenology emerged  as a method of work. 
Eventually, during MZ's master thesis  \cite{MZ}, %he could do
  a fine tuning  %("a secondary indicator" for the mentioned "operation") 
  of SVD was achieved and %bring forth
   the essential non-singular physically desirable results were brought forth for the first time in the fall of 2009. 

\end{document}